\documentclass[conference,a4paper]{APSIPA2021}
\usepackage{multirow}
\usepackage{graphicx}
\usepackage{amsmath}
\usepackage[psamsfonts]{amssymb}
\usepackage{amsxtra}
\usepackage{threeparttable}
\usepackage{booktabs}

\begin{document}

\title{Conformer-based End-to-end Speech Recognition With Rotary Position Embedding}

\author{%
\authorblockN{%
Shengqiang Li, Menglong Xu, Xiao-Lei Zhang
}
%
\authorblockA{%
CIAIC, School of Marine Science and Technology, Northwestern Polytechnical University, China\\
E-mail: \{shengqiangli,mlxu\}@mail.nwpu.edu.cn, xiaolei.zhang@nwpu.edu.cn}
}

\maketitle
\thispagestyle{empty}

\begin{abstract}
Transformer-based end-to-end speech recognition models have received considerable attention in recent years due to their high training speed and ability to model a long-range global context. Position embedding in the transformer architecture is indispensable because it provides supervision for dependency modeling between elements at different positions in the input sequence. To make use of the time order of the input sequence, many works inject some information about the relative or absolute position of the element into the input sequence. In this work, we investigate various position embedding methods in the convolution-augmented transformer (conformer) and adopt a novel implementation named rotary position embedding (RoPE). RoPE encodes absolute
positional information into the input sequence by a rotation matrix, and then naturally incorporates explicit relative position information into a self-attention module. To evaluate the effectiveness of the RoPE method, we conducted experiments on AISHELL-1 and LibriSpeech corpora. Results show that the conformer enhanced with RoPE achieves superior performance in the speech recognition task. Specifically, our model achieves a relative word error rate reduction of $8.70\%$ and $7.27\%$ over the conformer on ‘test-clean’ and ‘test-other’ sets of the LibriSpeech corpus respectively.
\end{abstract}

\section{Introduction}
The sequential order of an input sequence plays a vital role in many sequence learning tasks, particularly in speech recognition. Recurrent neural networks (RNNs) based models can learn the sequential order by recursively computing their hidden states along the time dimension. Convolutional neural networks (CNNs) based models can implicitly learn the position information of an input sequence by a padding operator \cite{islam2020much}. In recent years, transformer-based models have shown great superiority in various sequence learning tasks, such as machine translation \cite{vaswani2017attention}, language modeling \cite{radford2019language} and speech recognition \cite{dong2018speech}. The transformer-based models utilize a self-attention mechanism to model the dependency among different elements in the input sequence, which provides more efficient parallel computing than RNNs and can model longer context-dependency among elements than CNNs.

The transformer-based models dispense with recurrence, and instead rely solely on a self-attention mechanism to draw global dependencies among elements in the input sequence. However, the self-attention mechanism cannot model the sequential order inherently \cite{yun2019transformers}. There are various works injecting some information about the relative or absolute position of the elements of the input sequence into the transformer-based models.

One line of works focuses on absolute position embedding methods. The position embedding is added to the input embeddings usually. The original work \cite{vaswani2017attention} injected absolute position information to the input embeddings via a trigonometric position embedding. Specifically, the absolute position of each element in the input sequence is encoded into a vector, whose dimension is equal to the dimension of the input embeddings. Another work \cite{gehring2017convolutional} added absolute position information via the learned positional embedding instead of the pre-defined function, the learned position embedding can achieve competitive performance with the trigonometric position embedding. However, it cannot be extrapolated to a sequence length longer than the maximum sequence length of training utterances.

The other line of works focuses on relative position embedding, which typically inject relative position information into the attention calculation. Originally proposed by \cite{shaw2018self}, the relative position embedding method replaced absolute positions by taking into account the distance between sequence elements. It demonstrates significant improvement in two machine translation tasks. The method has also been generalized to language modeling \cite{dai2019transformer}, which helps the language model capture very long dependency between paragraphs. Some works also utilized relative position embedding to acoustic modeling in the speech recognition task \cite{gulati2020conformer,pham2020relative}, which help the self-attention module deals with different input lengths better than the absolute position embedding methods.

In addition to these approaches, \cite{wang2019encoding} proposed to model the position information in a complex space. \cite{liu2020learning} proposed to model the dependency of position embedding from the perspective of Neural ordinary differential equations \cite{chen2018neural}. \cite{su2021roformer} proposed to encode relative position by multiplying the context representations with a rotation matrix.

In this paper, we investigated various position embedding methods in the convolution-augmented transformer (conformer) for speech recognition. Motivated by \cite{su2021roformer}, we adopt a novel implementation named rotary position embedding (RoPE). RoPE formulates the relative position naturally by an inner product of the input vectors of the self-attention module in the conformer, right after the absolution position information being encoded through the rotation matrix. Experiments were conducted on AISHELL-1 and LibriSpeech corpora. Results show that the conformer enhanced with the RoPE performs superior over the original conformer. It achieves a character error rate of $4.69\%$ on the test set of AISHELL-1 dataset, and a word error rate of of $2.1\%$ and $5.1\%$ on the `test-clean' and `test-other' sets of LibriSpeech dataset respectively.

The remainder of this paper is organized as follows. Section \ref{models} describes the RoPE method and the architecture of our model. Section
\ref{experiments} presents experiments. Conclusion is given in Section \ref{conclusion}.

\section{Related Work}
The core module of transformer-based models is the self-attention module, assuming that $\boldsymbol{X} \in \mathbb{R}^{T \times d}$ denotes the input sequence, where $T$ is the sequence length, $d$ is the dimension, the self-attention module first incorporates the position information to the input sequence and transforms them into queries, keys, values vectors respectively:

\begin{equation}
	\begin{aligned}
		\boldsymbol{q}_{m} &=\operatorname{f}_q\left(\boldsymbol{x}_{m}, m\right) \\
		\boldsymbol{k}_{n} &=\operatorname{f}_k\left(\boldsymbol{x}_{n}, n\right) \\
		\boldsymbol{v}_{n} &=\operatorname{f}_v\left(\boldsymbol{x}_{n}, n\right)
	\end{aligned}
\end{equation}
where $\boldsymbol{q}_{m},\boldsymbol{k}_{n},\boldsymbol{v}_{n}$ incorporate $m$-th and $n$-th position information via the function $\operatorname{f}_q(\cdot),\operatorname{f}_k(\cdot),\operatorname{f}_v(\cdot)$ respectively. The attention weights calculated using the query and key vectors, and the output is the weighted sum of the value vector:

\begin{equation}
	\begin{aligned}
		a_{m, n} &=\operatorname{softmax} \left(\frac{\boldsymbol{q}_{m} \boldsymbol{k}_{n}^{\top}}{\sqrt{d}}\right) \\
		\mathbf{o}_{m} &=\sum_{n=1}^{T} a_{m, n} \boldsymbol{v}_{n}
	\end{aligned}
\end{equation}

\subsection{Absolute position embedding}

Assuming that $\boldsymbol{x}_m \in \mathbb{R}^{d}$ is the $m$-th element in the input sequence. The implementation of absolute position embedding can be formulated as:

\begin{equation}
	\begin{aligned}
		\operatorname{f}_q\left(\boldsymbol{x}_{m}, m\right) &= \left(\boldsymbol{x}_{m}+\boldsymbol{p}_{m}\right)\boldsymbol{W}_{q} \\
		\operatorname{f}_k\left(\boldsymbol{x}_{m}, n\right) &= \left(\boldsymbol{x}_{m}+\boldsymbol{p}_{m}\right)\boldsymbol{W}_{k} \\
		\operatorname{f}_v\left(\boldsymbol{x}_{m}, n\right) &= \left(\boldsymbol{x}_{m}+\boldsymbol{p}_{m}\right)\boldsymbol{W}_{v}
	\end{aligned}
	\label{eq_ape}
\end{equation}
where $\boldsymbol{W}_q, \boldsymbol{W}_k, \boldsymbol{W}_v \in \mathbb{R}^{d \times d_m}$ is the weight matrix of the linear projection layer of query, key and value vectors respectively, $d_m$ is the hidden size of the attention module, $\boldsymbol{p}_{m} \in \mathbb{R}^{d}$ is a vector depending of the position information of $\boldsymbol{x}_m$. In \cite{lan2019albert,clark2020electra}, $\boldsymbol{p}_{m} \in\left\{\boldsymbol{p}_{j}\right\}_{j=1}^{T}$ is a set of trainable vectors. Ref. \cite{vaswani2017attention} has proposed to generate $\boldsymbol{p}_m$ using the sinusoidal function:

\begin{equation}
	\begin{array}{ll}
		p_{m, 2 j}  &=\sin \left(m / 10000^{2 j / d}\right) \\
		p_{m, 2 j+1}  &=\cos \left(m / 10000^{2 j / d}\right)
	\end{array}
\end{equation}
where $m$ is the position and $j$ is the dimension.

\subsection{Relative position embedding}

In \cite{dai2019transformer}, the relative distance between elements in the input sequence was taken into account. Specifically, keeping the form of (\ref{eq_ape}), the term $\boldsymbol{q}_{m} \boldsymbol{k}_{n}^{\top}$ can be decomposed to:

\begin{equation}
	\begin{aligned}
		\boldsymbol{q}_{m} \boldsymbol{k}_{n}^{\top}&=\operatorname{f}_q(\boldsymbol{x}_m,m)\operatorname{f}_k^{\top}(\boldsymbol{x}_n,n)\\
		&=\boldsymbol{x}_{m} \boldsymbol{W}_{q} \boldsymbol{W}_{k}^{\top} \boldsymbol{x}_{n}^{\top}+\boldsymbol{x}_{m} \boldsymbol{W}_{q} \boldsymbol{W}_{k}^{\top} \boldsymbol{p}_{n}^{\top}\\
		&+\boldsymbol{p}_{m} \boldsymbol{W}_{q} \boldsymbol{W}_{k}^{\top} \boldsymbol{x}_{n}^{\top}+\boldsymbol{p}_{m} \boldsymbol{W}_{q} \boldsymbol{W}_{k}^{\top} \boldsymbol{p}_{n}^{\top}
	\end{aligned}
	\label{decomposition of qk}
\end{equation}
In \cite{dai2019transformer}, (\ref{decomposition of qk}) was modified to:

\begin{equation}
	\begin{aligned}
		\boldsymbol{q}_{m} \boldsymbol{k}_{n}^{\top}&=
		\boldsymbol{x}_{m} \boldsymbol{W}_{q} \boldsymbol{W}_{k,E}^{\top} \boldsymbol{x}_{n}^{\top}+\boldsymbol{x}_{m} \boldsymbol{W}_{q} \boldsymbol{W}_{k,R}^{\top} \boldsymbol{R}_{m-n}^{\top}
		\\
		&+\boldsymbol{u}\boldsymbol{W}_{k}^{\top} \boldsymbol{x}_{n}^{\top}+\boldsymbol{v} \boldsymbol{W}_{k}^{\top} \boldsymbol{R}_{m-n}^{\top}
	\end{aligned}
	\label{eq_rpe}
\end{equation}
where $\boldsymbol{u}, \boldsymbol{v}$ are trainable parameters, and $\boldsymbol{R}_{m-n}$ is the relative position embedding. Comparing the (\ref{eq_rpe}) and (\ref{decomposition of qk}), we can see that there are three main changes:
\begin{itemize}
	\item {Firstly, the absolute position embedding $\boldsymbol{p}_m$ for computing key representation is replaced with relative counterpart $\boldsymbol{R}_{m-n}$.}
	\item {Secondly, the query $\boldsymbol{p}_{m} \boldsymbol{W}_{q}$ is replaced by two trainable parameters $\boldsymbol{u}, \boldsymbol{v}$.}
	\item {Finally, the weight matrix of the linear projection layer of key vector is separated to two matrices $\boldsymbol{W}_{k,E}, \boldsymbol{W}_{k,R}$ for producing the content-based key vector and location-based key vector respectively.}
\end{itemize}

\section{Method}
\label{models}

In this section, we describe the rotary position embedding (RoPE) and illustrate how we apply it to the self-attention module in transformer-based models.

\subsection{Formulation}

Considering the dot-product attention does not preserve absolute positional information, so that if we encode the position information via absolute position embeddings, we will lose a significant amount of information. On the other hand, the dot-product attention does preserve relative position, so if we can encode the positional information into the input sequence in a way that only leverages relative positional information, that will be preserved by the attention function.

To incorporate relative position information, we hope the inner product encodes position information in the relative form only:
\begin{equation}
	\left\langle \operatorname{f}_{q}\left(\boldsymbol{x}_{m}, m\right), \operatorname{f}_{k}\left(\boldsymbol{x}_{n}, n\right)\right\rangle=\operatorname{g}\left(\boldsymbol{x}_{m}, \boldsymbol{x}_{n}, m-n\right)
	\label{formulation_rope}
\end{equation}
where the inner product of $\boldsymbol{q}_m$ and $\boldsymbol{k}_n$ is formulated by a function $\operatorname{g}(\cdot)$, which takes only $\boldsymbol{x}_m$, $\boldsymbol{x}_n$ and their relative position $m-n$ as input variables. Actually, finding such an encoding mechanism is equivalent to solving the function $\operatorname{f}_q(\cdot)$ and $\operatorname{f}_k(\cdot)$ that conform (\ref{formulation_rope}).

\subsection{Rotary position embedding}

We start with a simple case with dimension $d=2$, the RoPE provides a solution to (\ref{formulation_rope}):
\begin{equation}
	\begin{aligned}
		\operatorname{f}_{q}\left(\boldsymbol{x}_{m}, m\right) &=\left(\boldsymbol{x}_{m} \boldsymbol{W}_{q} \right) e^{i m \theta} \\
		\operatorname{f}_{k}\left(\boldsymbol{x}_{n}, n\right) &=\left( \boldsymbol{x}_{n} \boldsymbol{W}_{k}\right) e^{i n \theta} \\
		\operatorname{g}\left(\boldsymbol{x}_{m}, \boldsymbol{x}_{n}, m-n\right) &=\operatorname{Re}\left[\left( \boldsymbol{x}_{m} \boldsymbol{W}_{q}\right)\left( \boldsymbol{x}_{n} \boldsymbol{W}_{k} \right)^{*} e^{i(m-n) \theta}\right]
	\end{aligned}
\end{equation}
where $\operatorname{Re}[\cdot]$ denotes the real part of a complex number and $\left(\boldsymbol{x}_{n} \boldsymbol{W}_{k} \right)^{*}$ represents the conjugate complex number of $\left(\boldsymbol{x}_{n} \boldsymbol{W}_{k} \right)$, $\theta \in \mathbb{R}$ is a non-zero constant.

Considering the merit of the linearity of the inner product, we can generalize the solution to any dimension when $d$ is even, we divide the $d$-dimension space to $d/2$ sub-spaces and combine them:

\begin{equation}
	\begin{aligned}
		\operatorname{f}_{q}(\boldsymbol{x}_m, m)&=
		\boldsymbol{R}_{\Theta, m}^{d} \boldsymbol{x}_{m} \boldsymbol{W}_{q}\\		
		\operatorname{f}_{k}(\boldsymbol{x}_n, n)&=
		\boldsymbol{R}_{\Theta, n}^{d} \boldsymbol{x}_{n} \boldsymbol{W}_{k}
	\end{aligned}	
\end{equation}
where 	
\begin{equation}
	\boldsymbol{R}_{\Theta, m}^{d}=\left(\begin{array}{cccc}
		\boldsymbol{M}_{1} & & & \\
		& \boldsymbol{M}_{2} & & \\
		& & \ddots & \\
		& & & \boldsymbol{M}_{d / 2}
	\end{array}\right)
\end{equation}

\begin{equation}
	\Theta=\left\{\theta_{i}=10000^{-2(i-1) / d}, i \in[1,2, \ldots, d / 2]\right\}
\end{equation}

\begin{equation}
	\boldsymbol{M}_{i}=\left(\begin{array}{cc}
		\cos m \theta_{i} & -\sin m \theta_{i} \\
		\sin m \theta_{i} & \cos m \theta_{i}
	\end{array}\right)
\end{equation}

\begin{figure}[t]
	\begin{center}
		\includegraphics[width=9cm]{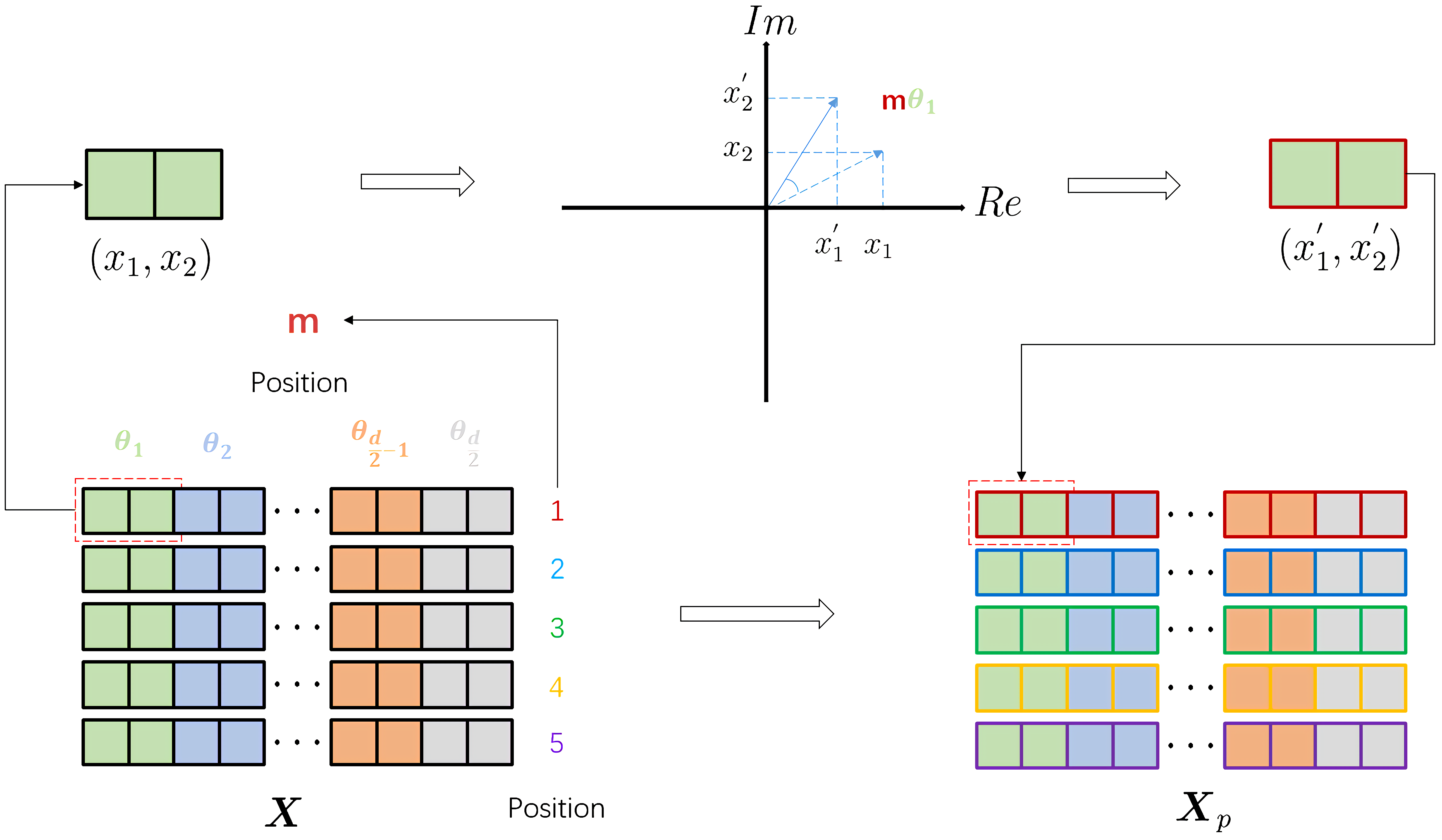}
	\end{center}
	\caption{Illustration of rotary position embedding (RoPE). $\boldsymbol{X}$ is the input sequence without position embedding and $\boldsymbol{X}_p$ is the sequence encoded with position information.}
	\vspace*{-3pt}
	\label{fig.RoPE}
\end{figure}

The illustration of rotary position embedding is shown in Figure \ref{fig.RoPE}.

\subsection{Enhanced conformer with RoPE}

\begin{figure}[t]
	\begin{center}
		\includegraphics[width=8cm]{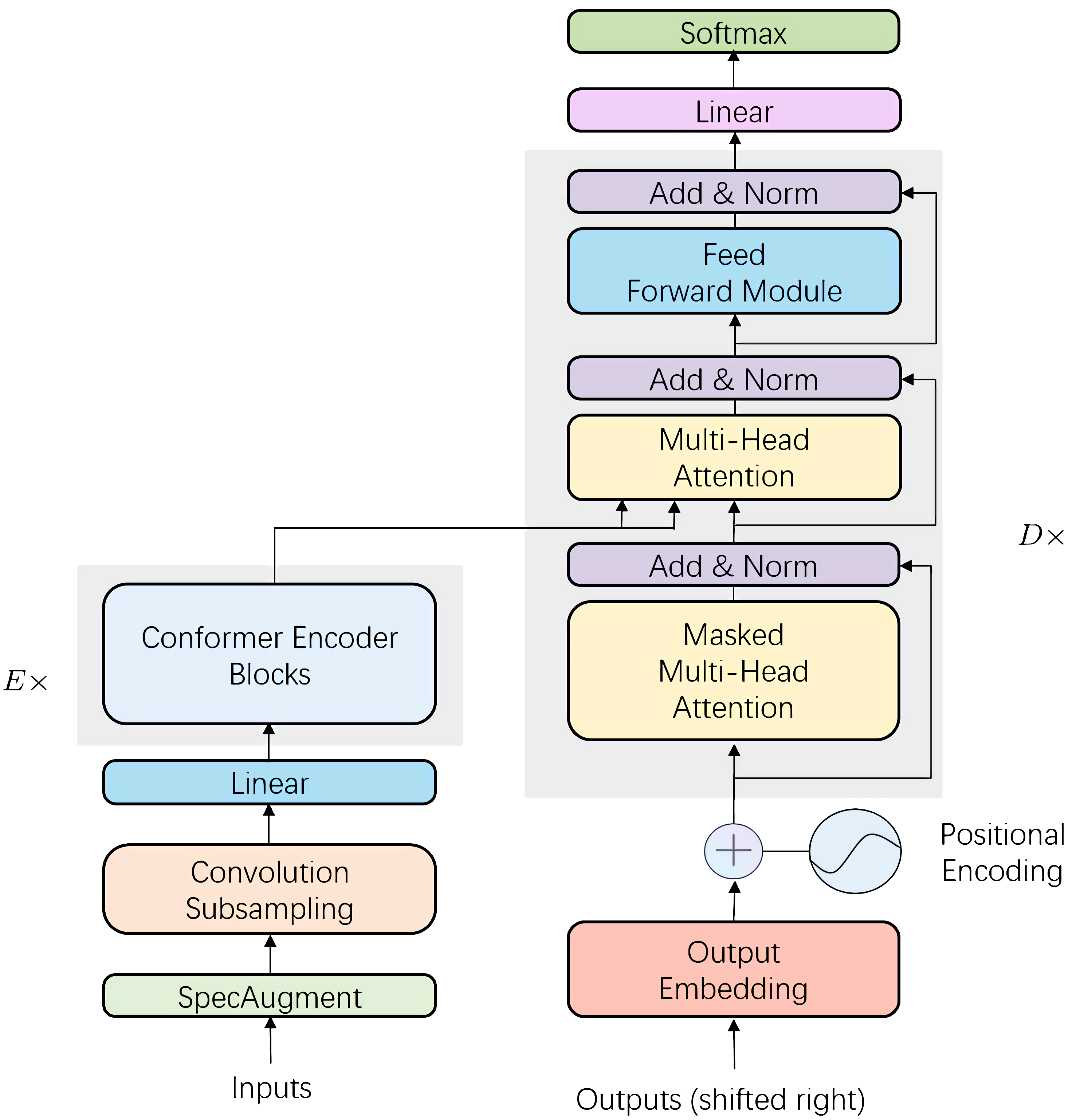}
	\end{center}
	\caption{The architecture of conformer.}
	\label{fig.Conformer}
\end{figure}

\begin{figure}[t]
	\begin{center}
		\includegraphics[width=8cm]{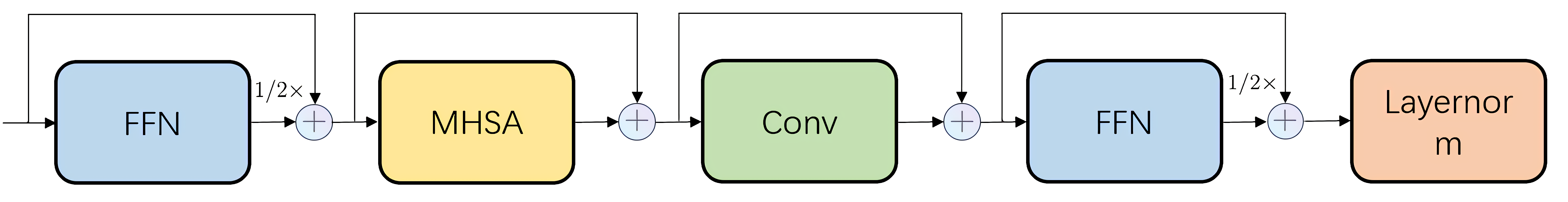}
	\end{center}
	\caption{The architecture of conformer encoder blocks.}
	\label{fig.Encoder}
\end{figure}

In this work, we adopt conformer \cite{gulati2020conformer} as the speech recognition model, which is a state-of-the-art transformer-based model. The architecture of the conformer is given in Figure \ref{fig.Conformer}. The audio encoder of conformer first processed the input with a convolution subsampling module and then with $E$ conformer encoder blocks. Each conformer encoder block contains two feed-forward (FFN) modules sandwiching the multi-head self-attention (MHSA) module and the convolution (Conv) module, as shown in Figure \ref{fig.Encoder}. Because the decoder of conformer is identical with transformer \cite{vaswani2017attention}, we will not describe the decoder anymore.

In contrast to the additive position embedding method used by other works \cite{vaswani2017attention}, we adopt the multiplicative position embedding method in the encoder. Moreover, we do not add the position embedding at the beginning of the encoder, but rather, we add the position embedding to the query and key vectors at each self-attention layer. The position embedding in the decoder is absolute position embedding, which is identical with the one in transformer \cite{vaswani2017attention}.

\section{Experiments}
\label{experiments}

\subsection{Datasets}

Our experiments were conducted on a Mandarin speech corpus AISHELL-1 \cite{bu2017aishell} and an English speech corpus LibriSpeech \cite{panayotov2015librispeech}. The former has 170 hours labeled speech, while the latter consists of 970 hours labeled speech and an additional 800M word token text-only corpus for building language model.

\subsection{Setup}

We used 80-channel log-mel filterbank coefficients (Fbank) features computed on a 25ms window with a 10ms shift. The features for each speaker were rescaled to have zero mean and unit variance. The token vocabulary of AISHELL-1 contains 4231 characters. We used a 5000 token vocabulary based on the byte pair encoding algorithm \cite{sennrich2015neural} for LibriSpeech. Moreover, the vocabularies of AISHELL-1 and LibriSpeech have a padding symbol '$\langle PAD\rangle$' , an unknown symbol '$\langle UNK\rangle$', and an end-of-sentence symbol '$\langle EOS\rangle$'.

Our model contains 12 encoder blocks and 6 decoder blocks. There are 4 heads in both the self-attention and the encoder-decoder attention. The 2D-CNN frontend utilizes two $3\times3$ convolution layers with 256 channels. The rectified linear units were used as the activation. The stride was set to 2. The hidden dimension of the attention layer is 256. The hidden dimension and output dimension of the feed-forward layer are 256 and 2048 respectively. We used the Adam optimizer and a transformer learning rate schedule \cite{vaswani2017attention} with 30000 warm-up steps and a peak learning rate of 0.0005. We used SpecAugment \cite{park2019specaugment} for data augmentation. We set the CTC weight to 0.3 for the joint training with the attention model. In the test stage, we set CTC weight to 0.6 for the joint decoding. We used a transformer-based language model to refine the results.

To evaluate the effectiveness of our model, we compare our model with 9 representative speech recognition models, which are TDNN-Chain (kaldi) \cite{povey2016purely}, LAS \cite{park2019specaugment}, SA-Transducer \cite{tian2019self}, Speech-Transformer \cite{tian2020spike}, LDSA \cite{xu2020transformer}, GSA-Transformer \cite{liang2021transformer}, Conformer \cite{guo2020recent}, Dynamic convolution (DC) \cite{fujita2020attention} and Self-attention dynamic convolution 2D (SA-DC2D) \cite{fujita2020attention}.
There are 4 state-of-the-art transformer-based models in the comparison methods. Speech-Transformer uses the transformer architecture for both acoustic modeling and language modeling. LDSA uses a local dense synthesizer attention module in the transformer encoder as an alternative of the self-attention module. GSA-Transformer replaces the self-attention module with a gaussion-based attention module. Conformer combines the transformer architecture with a convolution module.

\subsection{Main results}
\label{result}

Table \ref{tab.compare_a} lists the comparison result on LibriSpeech dataset. From the table, we can see that the proposed conformer enhanced with RoPE achieves the best performance among these methods. Our model achieves a WER of $2.1\%$ and $5.1\%$ on the on ‘test-clean’ and ‘test-other’ sets respectively, which gets a relative WER reduction of $8.70\%$ and $7.27\%$ over the conformer.

Table \ref{tab.compare_b} lists the comparison result on AISHELL-1 dataset. From the table, we see that the proposed model achieves a CER of $4.34\%$ on the development set and $4.69\%$ on the test set respectively, which gets a relative CER reduction of $4.00\%$ and $3.90\%$ on the development set and test set over the conformer. Moreover, the proposed model significantly outperforms the other comparison methods.

\begin{table}[t]
	\caption{ Comparison results on LibriSpeech. }
	\begin{center}
	\scalebox{1.0}{
		\begin{tabular}{lllll}
			\toprule
			
			\multirow{3}{*}{\textbf{Model}}  & \multicolumn{4}{c}{\textbf{WER(\%)}}\\
			& \multicolumn{2}{c}{\textbf{Dev}} & \multicolumn{2}{c}{\textbf{Test}} \\
			& \textbf{Clean} & \textbf{Other} & \textbf{Clean} & \textbf{Other} \\
			\midrule
			
			LAS \cite{park2019specaugment} & - & - & 2.5 & 5.8 \\
			
			DC \cite{fujita2020attention} & 3.5 & 10.5 & 3.6 & 10.8 \\
			
			SA-DC2D \cite{fujita2020attention} & 3.5 & 9.6 & 3.9 & 9.6 \\
			
			Conformer \cite{guo2020recent} & 2.1 & 5.5 & 2.3 & 5.5 \\
			
			Conformer (RoPE) & 1.9 & 5.0 & 2.1 & 5.1 \\
			
			\bottomrule
			
		\end{tabular}}
		\label{tab.compare_a}
	\end{center}
\end{table}

\begin{table}[t]
	\caption{Comparison results on AISHELL-1.}
	\begin{center}
		\scalebox{1.0}{
			\begin{tabular}{lll}
			\toprule
			
			\multirow{2}{*}{\textbf{Model}}  & \multicolumn{2}{c}{\textbf{CER(\%)}} \\
			& \textbf{Dev set} & \textbf{Test set} \\
			\midrule
			
			TDNN-Chain (kaldi) \cite{povey2016purely} & - & 7.45 \\
			
			SA-Transducer \cite{tian2019self} & 8.30 & 9.30 \\
			
			Speech-Transformer \cite{tian2020spike}  & 6.57 & 7.37 \\
			
			LDSA \cite{xu2020transformer} & 5.79 & 6.49 \\
			
			GSA-Transformer \cite{liang2021transformer} & 5.41 & 5.94 \\
			
			Conformer \cite{guo2020recent} & 4.52 & 4.88\\
			
			Conformer (RoPE) & 4.34 & 4.69 \\
			
			\bottomrule
			
		\end{tabular}
	}

		\label{tab.compare_b}
	\end{center}
\end{table}

\subsection{Comparison of different position embedding methods}
\label{comparison}

We also compare the rotary position embedding with other position embedding in the conformer architecture, i.e. absolute position embedding and relative position embedding. Table \ref{tab.compare_c} lists the result on LibriSpeech dataset and Table \ref{tab.compare_d} lists the result on AISHELL-1. From Table \ref{tab.compare_c} and Table \ref{tab.compare_d}, we can see that the relative position embedding performs better than the absolute position embedding, and the rotary position embedding achieves the best performance among these position embedding methods on both LibriSpeech and AISHELL-1 dataset.

\begin{table}[t]
	\caption{Comparison between position embedding methods on the LibriSpeech dataset. APE denotes absolute position embedding, RPE denotes relative position embedding respectively.}
	\begin{center}
		\scalebox{1.0}{
		\begin{tabular}{lllll}
			\toprule
			\multirow{3}{*}{\textbf{Model}}  & \multicolumn{4}{c}{\textbf{WER(\%)}}\\
			& \multicolumn{2}{c}{\textbf{Dev}} & \multicolumn{2}{c}{\textbf{Test}} \\
			& \textbf{Clean} & \textbf{Other} & \textbf{Clean} & \textbf{Other} \\
			\midrule
			
			Conformer (APE) & 2.1 & 5.5 & 2.3 & 5.5 \\
			
			Conformer (RPE) & 2.0 & 5.2 & 2.2 &  5.5 \\
			
			Conformer (RoPE) & 1.9 & 5.0 & 2.1 & 5.1 \\
			\bottomrule			
		\end{tabular}	
}

		\label{tab.compare_c}
	\end{center}
\end{table}

\begin{table}[htbp]
	\caption{Comparison between position embedding methods on the AISHELL-1 dataset.}
	\begin{center}
		\scalebox{1.0}{
		\begin{tabular}{lll}
			\toprule
			\multirow{2}{*}{\textbf{Model}}  & \multicolumn{2}{c}{\textbf{CER(\%)}} \\
			& \textbf{Dev set} & \textbf{Test set} \\
			\midrule
			
			Conformer (APE)  & 4.52 & 4.88 \\
			
			Conformer (RPE) & 4.49 & 4.82\\
			
			Conformer (RoPE) & 4.34 & 4.69 \\
			
			\bottomrule
		\end{tabular}
	
}

	\label{tab.compare_d}
	\end{center}
\end{table}

\section{Conclusions}
\label{conclusion}

Transformer-based models have received great popularity in the speech recognition task. Position embedding of the input sequence plays a significant role in transformer-based models. In this paper, we propose to apply the rotary position embedding into the conformer. The rotary position embedding incorporates explicit relative position information in the self-attention module to enhance the performance of the conformer architecture. Our experimental results on the AISHELL-1 and LibriSpeech corpora demonstrate that the enhanced conformer with rotary position embedding performs superior over the vanilla conformer and several representative models.

\bibliographystyle{IEEEtran}
\bibliography{refs}

\end{document}